\documentclass[10pt]{iopart}

\usepackage{graphicx} 
\usepackage{lineno,hyperref}
\hypersetup{colorlinks,linkcolor={red!50!black},citecolor={green!50!black}, urlcolor={blue!50!black}}
\modulolinenumbers[5]
\usepackage{xcolor}
\usepackage{amssymb}
\usepackage{acro}
\DeclareAcronym{2PM}{short=2PM,long=2-Point Model}
\DeclareAcronym{3LP}{short=3LP,long=triple-tipped Langmuir probe}
\DeclareAcronym{Afrac}{short=A$_{frac}$\ , long=attachment fraction}
\DeclareAcronym{afrac}{
    short=$J_{sat}/J_{roll}$,
    long=\ac{Jsat} normalized to its value at rollover
}
\DeclareAcronym{CLTF}{short=CLTF,long=closed loop transfer function}
\DeclareAcronym{DTS}{short=DTS, long=Divertor Thomson Scattering}
\DeclareAcronym{FOPDT}{short=FOPDT,long=First Order Plus Dead Time}
\DeclareAcronym{GGD}{short=GGD, long=Generalized Grid Description}
\DeclareAcronym{IDS}{short=IDS, long=Interface Data Structure}
\DeclareAcronym{IMAS}{short=IMAS, long=ITER Integrated Modelling and Analysis Suite}
\DeclareAcronym{IRVB}{short=IRVB, long=Infra-Red Video Bolometer}
\DeclareAcronym{ITB}{short=ITB, long=Internal Transport Barrier}
\DeclareAcronym{ITER}{short=ITER, long=International Thermonuclear Experimental Reactor}
\DeclareAcronym{Jsat}{short=$J_{sat}$,long=ion saturation current density}
\DeclareAcronym{LCFS}{short=LCFS, long={Last Closed Flux Surface}}
\DeclareAcronym{LP}{short=LP,long=Langmuir probe}
\DeclareAcronym{MIMO}{short=MIMO, long={multi in, multi out}}
\DeclareAcronym{MPC}{short=MPC, long=Model Predictive Control}
\DeclareAcronym{NBI}{short=NBI, long=neutral beam injection}
\DeclareAcronym{OLTF}{short=OLTF,long=open loop transfer function}
\DeclareAcronym{OSP}{short=OSP, long=outer strike point}
\DeclareAcronym{PCS}{short=PCS,long=plasma control system}
\DeclareAcronym{PFR}{short=PFR,long=private flux region}
\DeclareAcronym{PID}{short=PID,long=proportional-integral-derivative}
\DeclareAcronym{Prad}{short=$P_{rad}$,long=radiated power}
\DeclareAcronym{Pradcore}{short=$P_{rad,core}$,long=core radiated power}
\DeclareAcronym{PSOL}{short=$P_{SOL}$, long=power crossing into the scrape-off layer}
\DeclareAcronym{PVLC}{short=PVLC, long=Predicted Variable Linear Controller}
\DeclareAcronym{RT}{short=RT, long=real-time}
\DeclareAcronym{SISO}{short=SISO, long={single in, single out}}
\DeclareAcronym{SOPDT}{short=SOPDT,long=Second Order Plus Dead Time}
\DeclareAcronym{SOL}{short=SOL, long=Scrape-off Layer}
\DeclareAcronym{SOLPS-ITER}{short=SOLPS-ITER, long=Scrape-off Layer Plasma Simulator - ITER version}
\DeclareAcronym{UGF}{short=UGF, long=unity gain frequency}
\DeclareAcronym{ZN}{short=Z-N, long=Ziegler-Nichols}
\DeclareAcronym{FPP}{short=FPP, long=Fusion Pilot Plant}

\newcommand{\Afrac}{A$_{frac}$\ }

\def\GA{General Atomics, 3550 General Atomics Ct., San Diego, CA 92121, USA}
\def\ORAU{Oak Ridge Associated Universities, 100 ORAU Way, Oak Ridge, TN 37830, USA}

\def\LLNL{Lawrence Livermore National Laboratory, PO Box 808, Livermore, CA 94550, USA}

\def\KFE{Korea Institute of Fusion Energy, 169-148 Gwahak-ro, Yuseong-gu, Daejeon 34133, Republic of Korea}
\def\KAIST{Korea Advanced Institute of Science and Technology, 291 Daehak-ro, Yuseong-gu, Daejeon 34141, Republic of Korea}

\begin{document}

\title{Detachment control in KSTAR with Tungsten divertor}

\author{
    Anchal Gupta$^{a, b}$\footnote{Corresponding author: guptaa@fusion.gat.com},
    David Eldon$^b$, 
    Eunnam Bang$^c$, 
    KyuBeen Kwon$^{a, b}$, 
    Hyungho Lee$^c$, 
    Anthony Leonard$^b$, 
    Junghoo Hwang$^{c, d}$, 
    Xueqiao Xu$^e$, 
    Menglong Zhao$^e$, 
    Ben Zhu$^e$
}

\address{$^a$\ORAU}
\address{$^b$\GA}
\address{$^c$\KFE}
\address{$^d$\KAIST}
\address{$^e$\LLNL}

\date{\today}

\begin{abstract}
KSTAR has recently undergone an upgrade to use a new Tungsten divertor to run experiments in ITER-relevant scenarios.
Even with a high melting point of Tungsten, it is important to control the heat flux impinging on tungsten divertor targets to minimize sputtering and contamination of the core plasma.
Heat flux on the divertor is often controlled by increasing the detachment of \ac{SOL} plasma from the target plates.
In this work, we have demonstrated successful detachment control experiments using two different methods.
The first method uses attachment fraction as a control variable which is estimated using ion saturation current measurements from embedded Langmuir probes in the divertor.
The second method uses a novel machine-learning-based surrogate model of 2D UEDGE simulation database, DivControlNN.
We demonstrated running inference operation of DivControlNN in realtime to estimate heat flux at the divertor and use it to feedback impurity gas to control the detachment level.
We present interesting insights from these experiments including a systematic approach to tuning controllers and discuss future improvements in the control infrastructure and control variables for future burning plasma experiments.
\end{abstract}

\submitto{\PPCF}

\maketitle

\ioptwocol

\acresetall  

\section{Introduction}
\label{sec:introduction}

Burning plasma tokamaks such as ITER\cite{Holtkamp_2007_FED}, SPARC\cite{Creely_2020_JPP}, and the various DEMO\cite{Federici_2014_FED} and \ac{FPP}\cite{Buttery_2021_NF} concepts are estimated to exhaust very high heat flux in the \ac{SOL} towards the divertor target.
Heat exhausted from the core plasma is rapidly conducted along the open field lines in the \ac{SOL} toward the divertor targets.
To withstand the high heat flux, the divertor target plates are planned to be made out of tungsten, which has a high melting point, good resilience against erosion by the plasma, and relatively low tritium retention compared to other well-studied materials like carbon.
Using ITER's tungsten divertor as an example, it is estimated that for steady-state operation, the constant heat flux reaching the divertor target has to be below 10-15 MW/m$^2$\cite{Pitts_2019_NME} to avoid surface melting and structural damage to the divertor plates.
Additionally, tungsten being a very high-Z material poses contamination challenges for the core plasma and it is important to develop operation strategies that limit the tungsten sputtering, especially in the divertor region where hot plasma interacts with the tungsten surface in a very narrow region of the order of a few mm\cite{Eich_2013_NF}.
Therefore, the electron temperature at the target plate must be below the tungsten sputtering threshold, which is 8 eV\cite{Brezinsek_2019_NF} for sputtering by deuterium but lower for heavier ions, to minimize contamination of the core plasma.
Thus, experimental reactors such as KSTAR are in the process of changing their divertor and main chamber walls from carbon to tungsten to facilitate study of plasma scenarios, operations, and control in the presence of a reactor-relevant wall.

The heat flux reaching the divertor as well as $T_e$ near the target are both typically reduced by puffing in gas in the \ac{SOL} region to dissipate energy and momentum from the exhaust plasma through ionization, charge exchange, and radiation.
As the puffed gas neutrals travel toward the core plasma, they radiate energy based on local electron temperature and density.
Hydrogenic fuel and helium exhaust atoms are ionized at low temperatures in the \ac{SOL} and divertor, and thus offer little radiative cooling except at the edges where the plasma has already cooled.
Thus for effective and fast cooling, impurity gases such as nitrogen, neon, and argon are puffed which can dissipate heat through radiation farther away from the divertor.
In the presence of such radiative dissipation, the heat load conducted by the plasma to the divertor reduces.
Additionally, high density (facilitated both by adding gas and by cooling the plasma at $\approx$constant pressure) dissipates momentum and reduces the total ion flux which impinges on the divertor.
When these dissipation process become significant, recombination occurs and the divertor begins to be shielded from the plasma by a population of neutrals: the primary plasma-neutral interaction zone \emph{detaches} from the solid target plate.
When only part of the surface is detached, the plasma is said to be partially detached, while if the ion flux is almost completely stopped with higher neutral gas pressure, it is said to be fully detached.

It is important though to keep the amount of impurity gases injected into the vessel to a minimum as higher gas injection eventually leads to more impurity reaching in the pedestal region of the plasma.
This can lead to rapid cooling which can collapse H-mode and could also lead to disruption of the plasma confinement.
Such sudden loss of plasma confinement can cause damage to the plasma-facing components.
Thus, it is important to carefully control the amount of impurity injected to keep the divertor cool while not contaminating the core plasma too much.

There are two key approaches to controlling divertor conditions.
The first is to control the sources that determine the fluxes that reach the divertor, such as radiated power throughout the plasma, sometimes divided between the divertor volume, \ac{SOL}, and within the \ac{LCFS}.
This has been successfully demonstrated in various machines:
using the bolometer chords in divertor region in Alcator C-Mod\cite{Goetz_1999_POP}, JT-60U\cite{Asakura_2009_NF}, ASDEX Upgrade\cite{Kallenbach_2012_NF} and DIII-D\cite{Eldon_2019_NME},
using AXUV diodes in EAST\cite{Wu_2018_NF},
using VUV N VII line emission in JET\cite{Maddison_2011_NF}, and
using C-III emission radiation front measured using MANTIS in TCV\cite{Ravensbergen_2021_NC}.

The second way is to directly control key parameters at the divertor target plates.
This has been demonstrated widely in several machines as well:
using divertor plate temperature measurements with surface thermocouples in Alcator C-Mod\cite{Brunner_2017_NF},
using surface electron temperature measurements with triple-tip Langmuir probes in EAST\cite{Eldon_2021_NME} or $T_e$ very close to the surface with divertor Thomson scattering in DIII-D \cite{Eldon_2017_NF},
using ion saturation current measurements from embedded Langmuir probes in JET\cite{Guillemaut_2017_PPCF}, EAST\cite{Yuan_2020_FED}, DIII-D\cite{Eldon_2021_NME}, and COMPASS\cite{Khodunov_2021_PPCF}.
In KSTAR, the ion saturation current measurements along with core electron density, injected power, and local magnetic field were used to calculate a derived control variable, \ac{Afrac}, which was used to control the detachment\cite{Eldon_2022_PPCF}.
This technique has the added benefit that the rollover ion saturation current does not need to be calculated or estimated prior to the shot and thus this technique has the potential for wider applicability in different scenarios and other machines.
In this work, we have re-used this technique in our experiments at KSTAR with a tungsten divertor to test the robustness of this control variable in the presence of high-Z contamination from tungsten.

In this work, we have also tested a new technique that uses a machine-learning-based surrogate model, DivControlNN\cite{Zhu_2025_InPrep}.
This model integrates measurements from several realtime inputs to run through a large database of 2D UEDGE\cite{Rognlien_1999_PP} simulations and provide a realtime estimate of the heat flux reaching the divertor plates along with several other key plasma parameters upstream in \ac{SOL} and at the two divertors.
We tested a prototype of this model with training and input limitations in KSTAR and demonstrated detachment control for the first time using such a surrogate model.
This paves the way for utilizing such models in future reactors that will have a very limited set of sensors available for control systems.

This paper is organized as follows.
In Sec.\ref{sec:control_variables}, we describe the experimental setup and the definition of different control variables used for detachment control.
In Sec.\ref{sec:sysid}, we describe our experimental shots used for identifying the system and using the fitted plant model to tune a PI controller using frequency response for closed-loop stability analysis and optimization.
In Sec.\ref{sec:results}, we show the results of our detachment control attempts.
Finally, in Sec.\ref{sec:discussion}, we discuss our results, the possible control and technical improvements we can make in the future, and how these results can aid in designing further experiments on KSTAR and other tokamaks.

\begin{figure}[!ht]
 \centering
 \includegraphics[width=\linewidth]{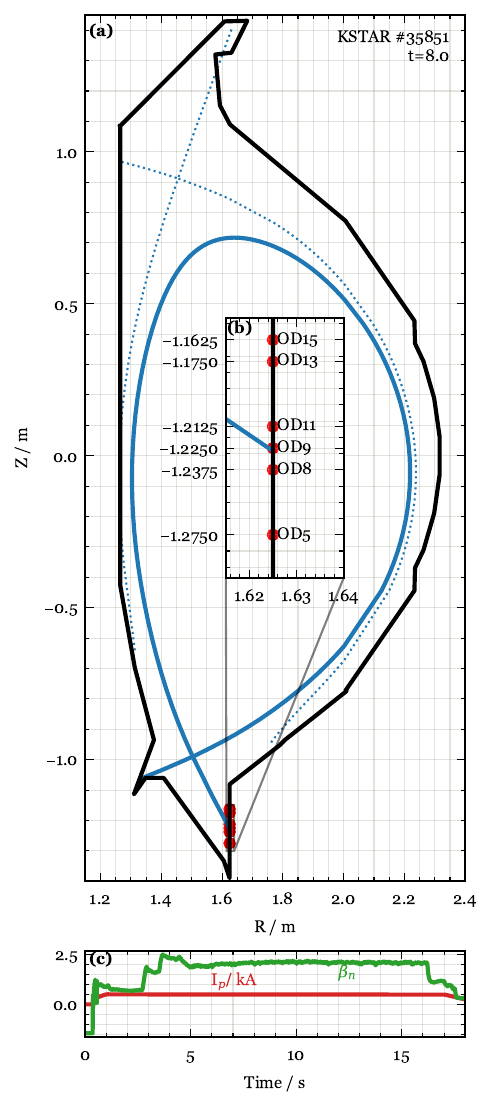}
 \caption{
Reference shot \#35851.
(a) Showing last closed flux surface (solid blue) and the secondary separatrix (dotted blue) at t=8 seconds.
The magnetic shape control was programmed to keep X point fixed which provided a sufficiently stable strike point on the realtime Langmuir Probe array.
(b) Zoomed-in locations of realtime Outer Divertor (OD) Langmuir probes.
(c) Plasma current (I$_p$) and $\beta_n$ for reference shot.
}
 \label{fig:ref_shot}
\end{figure}

\section{Experimental setup and control variables}
\label{sec:control_variables}

\begin{figure}[!ht]
 \centering
 \includegraphics[width=\linewidth]{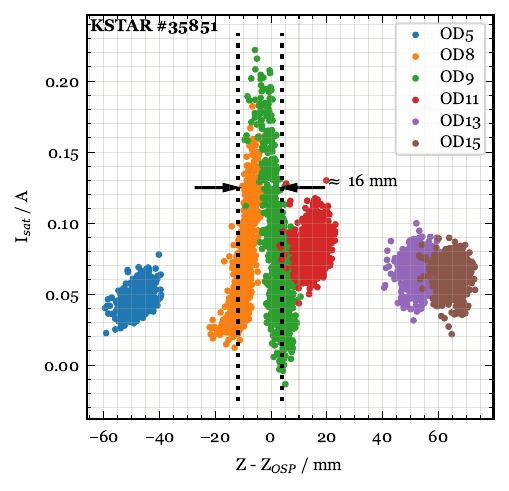}
 \caption{
Strike point width estimation for reference shot \#35851.
The raw data from langmuir probe array has been filtered by 4th order Butterworth filter with cut-off frequency of 50 Hz and then down sampled to 100 Hz.
For each data point on this plot, the x-axis position is calculated by subtracting the \ac{OSP} position reported by EFIT from the probe's Z coordinate.
The black dotted lines represent the rough estimate for width of strike point ion saturation current profile taken at half the maximum value above baseline.
}
 \label{fig:strike_point_width}
\end{figure}

The experiment was conducted on a standard lower single null H-mode plasma profile with reference shot KSTAR \#35851 with the equilibrium profile as shown in Fig.\ref{fig:ref_shot}.
The plasma shaping steps commenced by 7~s and the shot was programmed for flat-top up to 17~s providing a 10~s long window for the detachment control experiment.
For heat flux control, N$_2$ gas puffing was used.
The heat flux control variable was tested with several different inputs.

First, we utilized previously developed \ac{Afrac}\cite{Eldon_2022_PPCF}, which is defined as the ratio of measured ion saturation current ($I_{sat, measured}$) to modeled (using 2PM\cite{Leonard_2018_PPCF}) ion saturation current assuming fully attachment plasma ($I_{sat, attached}$).
\begin{equation}
    A_{frac} = \frac{I_{sat, measured}}{I_{sat, attached}}
\end{equation}

$I_{sat, attached}$ is estimated using Eq.(13) from \cite{Eldon_2022_PPCF}:

\begin{equation}
    I_{sat, attached} = C \langle n_{e} \rangle^2 q_{||, a}^{-\frac{3}{7}} 
\end{equation}

Here, $C$ is a calibration constant determined during reference shots so that \ac{Afrac} is 1.0 when \ac{SOL} plasma is fully attached to the divertor, $\langle n_{e} \rangle$ is the line-averaged electron density measured by interferometer and $q_{||, a}$ is the heat flux density at the outer mid-plane which is estimated using Eq.(15) from \cite{Eldon_2022_PPCF}.
The calibration constant $C$ accounts for gaps in real-time data availability on KSTAR and may be removed if more measurements become available in the future.
\ac{Afrac} is a convenient choice of control variable that is easily available in most tokamaks and allows for cross-comparison among machines.
If the strike point on the divertor tile is fixed in position well enough by the shape control system, a single close-by Langmuir probe is enough to provide the ion saturation current required for \ac{Afrac} calculation.
However, if the strike point control is not good enough, or if it is required to leave it as a free variable to allow for controlling other parameters in the shape control loop (as was the case in our experiments), then it is required to estimate the true ion saturation current through measurements made by a Langmuir probe array.
In our experiments, we chose the peak value from the Langmuir probe array as the input to the ion saturation current at the strike point.
Fig.\ref{fig:strike_point_width} plots the data from this Langmuir probe array for our reference shot.
The horizontal axis in this figure has been referenced from the EFIT reported \ac{OSP} position.
Thus, this figure shows the spread of the ion saturation current profile across the strike point.
Here, we see that the strike point is closer to OD8 and OD9 with a peak ion saturation current value of roughly 0.2~A at the strike point position.
We estimate the width of the ion saturation current profile at half the maximum value referenced to the baseline value of 0.05~A measured by far away probes, giving FWHM$\gtrsim$16~mm.
This ensures that when the strike point is within the closely placed probes, OD8, OD9, and OD11 (Fig.\ref{fig:ref_shot}), at least one probe can measure the ion saturation current while being within the peak region of the strike point.
We used the maximum value measured among the probe array to calculate \ac{Afrac} and since these probes are 12.5~mm apart, it means that the maximum deviation from the actual peak value would be $\lesssim$35\%.
Assuming that the strike point stays for equal amount of time in any location between the probe array (uniformly distributed, this can be further corraborated by noticing the motion of strike point in the figures in later sections), the mean error in peak value would be about 13\% while median error would be about 10\% assuming a gaussian profile with 16~mm FWHM.
This estimation in turn sets goals for a potential future strike point controller, to bound the strike point movement within 6.25~mm of the probe location to achieve above mentioned uncertainties.
If such a strike point controller can keep the strike point motion within 2.35~mm, the mean error would go below 2\% which would already be better than the other sources of error in the ion saturation current measurement.

Langmuir probes would not be able to survive high heat flux in burning plasma future reactors.
In general, such reactors would be severely limited in the number of realtime sensors available for control systems because of high neutron fluence and heat flux in vacuum vessels, and thus alternate control variables need to be searched for.
Toward this goal, we tested a prototype of a machine-learning-based surrogate model of 2D UEDGE, DivControlNN.
The employed version of DivControlNN is trained on approximately 70,000 2D UEDGE simulations of KSTAR.
The training dataset scanned core electron density ($1.5 \times 10^{19} - 7.0 \times 10^{19}$ m$^{-3}$), plasma current ($600-800$ kA), injected total power through NBI and ECH ($1-8$ MW), impurity fraction with respect to Deuterium density ($0-0.04$), and scaling of diffusion coefficient profile with a factor ($0.6 - 2$).
The diffusion coefficient profile is assumed for a typical H-mode shot which can be scaled as an input to the model.
This provided a widely applicable surrogate model that gives steady-state values of heat flux, ion saturation current, and electron temperature along the two divertors, electron density and temperature at the upstream point of the midplane, and total radiated power, power fraction radiated from divertor, and peak radiation power location in the poloidal cross-section of the device.
The model generates output within 20\% error from the 2D UEDGE output.

DivControlNN was originally developed and trained using Python's TensorFlow package and consists of two different models working in tandem.
The first is a multi-modal $\beta$-variational autoencoder\cite{Higgins_2017_ICLR} model to compress various quantities of interest coming from synthetic diagnostics on a 2D UEDGE database into a latent space representation.
The second stage is a multi-layer perceptron (MLP) model that maps the inputs of the 2D UEDGE simulations (which also form the inputs to the overall surrogate model).
During inference operation, the MLP model first maps the inputs to the latent space and the decoder network from the autoencoder then decodes the latent space into useful outputs.
While Python is the industry choice for developing and training such models, it can not be used for real-time inference purposes such as our use case.
We converted the Python model into a pure C code using a keras2c\cite{keras2c} package which is developed for generally converting such neural networks into real-time compatible C codes.
The generated C code runs an inference operation in about 160 $\mu$s on Intel\textsuperscript{\textregistered} Core\textsuperscript{TM} i7-6600U CPU @ 2.60GHz while we saw speed up of up to 18 $\mu$s per inference on Apple\textsuperscript{\textregistered} M2\textsuperscript{TM} Pro.
The real-time PCS in KSTAR runs its divertor control categories in a 50 $\mu$s clock cycle CPU, so we ran DivControlNN in a separate 1 ms clock-cycle CPU ensuring enough runtime for it along with other processes in that CPU.
This was still more than sufficient for our control purposes which anyway can not control faster than a few 10s of Hz due to system response time and gas actuation speed.

This preliminary model, however, has been trained on 2D UEDGE simulations of KSTAR with carbon divertor and carbon as the sole impurity species.
So the model does not exactly capture the environment with tungsten impurity from the tungsten divertor it was tested in, however the radiation loss profiles due to carbon and nitrogen (which was used in the test) are sufficiently similar to expect ballpark accuracy at the very least.
So it does not reflect the same Tungsten divertor system in which it was tested.
There were several other limitations to the realtime input provided to the model.
There was no reliable input for impurity fraction in plasma and we created an ad-hoc gas accumulation model which estimated impurity fraction by taking the ratio of total puffed impurity with total puffed Deuterium gas with estimated decay rates to model the effect of pumping and wall adsorption.
It turned out that even this estimator did not work correctly during the shots and we discuss this more in Sec.\ref{sec:sysid} later.
Additionally at KSTAR, the total input power from NBI and ECH sources is not completely available in realtime PCS and we had to input a feedforward signal matching the programmed rate of some sources that got summed with the other sources whose power was available in realtime.
Such feedforward programming is vulnerable to changes in actal power delivered during the shot including timing mismatch of on/off commands of NBI sources as well as power drop out when a source fails during the shot.
The ohmic power contribution is also prone to errors as a simple production $P_{ohm} \approx I_p \cdot V_{loop}$ is used in realtime estimate which assumes that all $I_p$ is inductively driven and ignores current drive due to other sources.
Finally, the diffusion coefficient scaling factor was set to 1.0 for lack of any better realtime information on it.
Despite these limitations, we attempted to use this model as a preliminary test for using such a surrogate model in real time and identify major obstacles before testing an improved and more relevant version in the future.

\newcommand{\AfracK}{K = -0.275$\pm$0.002}
\newcommand{\AfracTau}{$\tau$ = 1.00$\pm$0.02s}
\newcommand{\AfracL}{L = 0.154$\pm$0.006s}

\newcommand{\AfracKp}{K$_p$ = -10.0}
\newcommand{\AfracTi}{T$_i$ = 253.0 ms}
\newcommand{\Afracstau}{$\tau_s$ = 50.0 ms}
\newcommand{\AfracUGF}{0.59 Hz}
\newcommand{\AfracPhaseMargin}{14.8 $^\circ$}
\newcommand{\AfracDelayMargin}{69 ms}

\newcommand{\SMK}{K = -0.302$\pm$0.004}
\newcommand{\SMTau}{$\tau$ = 0.31$\pm$0.03s}
\newcommand{\SML}{L = 0.536$\pm$0.019s}

\newcommand{\SMKp}{K$_p$ = -3.0}
\newcommand{\SMTi}{T$_i$ = 68.5 ms}
\newcommand{\SMstau}{$\tau_s$ = 5.0 ms}
\newcommand{\SMUGF}{1.01 Hz}
\newcommand{\SMPhaseMargin}{11.9 $^\circ$}
\newcommand{\SMDelayMargin}{33 ms}

\begin{figure*}[!ht]
 \centering
 \includegraphics[width=\textwidth]{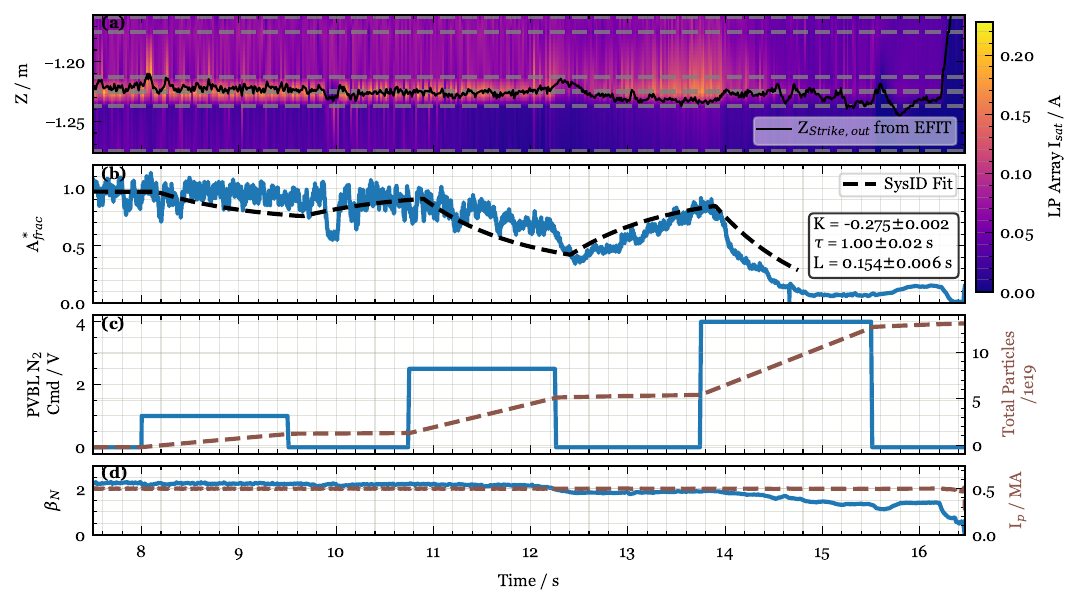}
 \caption{System identification shot \#35853.
(a) Shows the measured ion saturation current by realtime Langmuir Probe array at locations marked by grey dashed lines.
The data has been interpolated spatially using cubic spline interpolation.
The black curve shows the post-shot calculated strike point position on outer divertor using EFIT.
(b) Shows the \Afrac calculated from peak value among the Langmuir probe array.
The dashed black line shows the system identification fit on this data.
(c) Left axis: Shows the N$_2$ gas command steps sent for system identification.
Right axis: Shows the cummulative N$_2$ gas particles injected into the vessel.
(d) Left axis: Shows $\beta_n$.
Right axis: Shows the plasma current (I$_p$).
$^*$ Note: \Afrac for this shot was not calibrated properly and the raw data reported 2 times the value.
We fixed this factor after this shot and this figure shows the corrected value.
}
 \label{fig:sysid_afrac}
\end{figure*}

\begin{figure}[!ht]
 \centering
 \includegraphics[width=\linewidth]{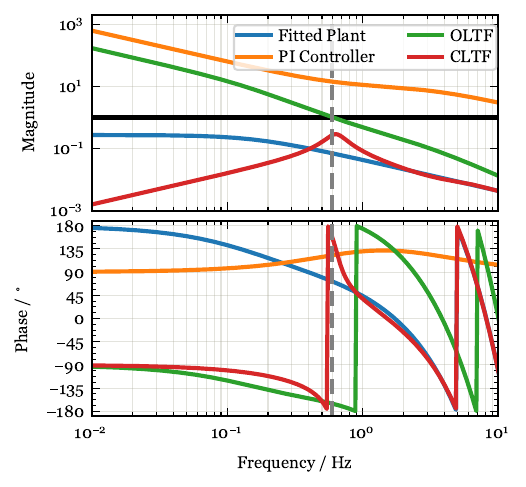}
 \caption{
Closed loop transfer function analysis of the system  using \Afrac output with chosen PI controller with gains:\AfracKp, \AfracTi, and \Afracstau.
The dashed grey vertical line shows the \ac{UGF} of the system.}
\label{fig:cltf_afrac}
\end{figure}

\section{System identification and Controller Tuning}
\label{sec:sysid}

\begin{figure*}[!ht]
 \centering
 \includegraphics[width=\textwidth]{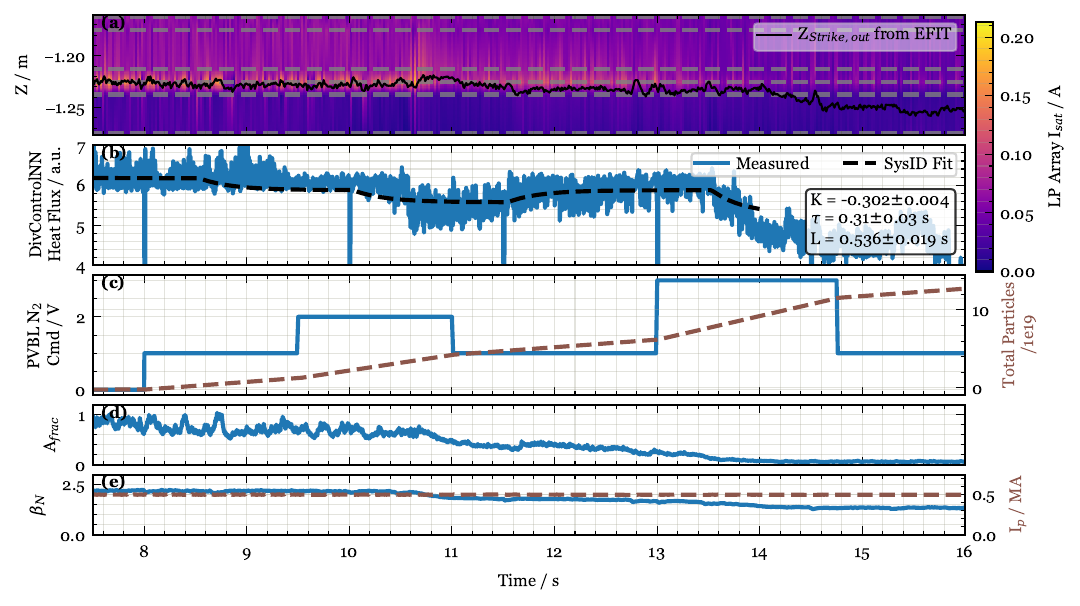}
 \caption{System identification shot \#35854.
(a) Shows the measured ion saturation current by realtime Langmuir Probe array at locations marked by grey dashed lines.
The data has been interpolated spatially using cubic spline interpolation.
The black curve shows the post-shot calculated strike point position on outer divertor using EFIT.
(b) Shows the heat flux at outer divertor calculated by DivControlNN.
The dashed black line shows the system identification fit on this data.
(c) Left axis: Shows the N$_2$ gas command steps sent for system identification.
Right axis: Shows the cummulative N$_2$ gas particles injected into the vessel.
(d) Shows the \Afrac calculated from peak value among the Langmuir probe array.
(e) Left axis: Shows $\beta_n$.
Right axis: Shows the plasma current (I$_p$).}
\label{fig:sysid_sm}
\end{figure*}

\begin{figure}[!ht]
 \centering
 \includegraphics[width=\linewidth]{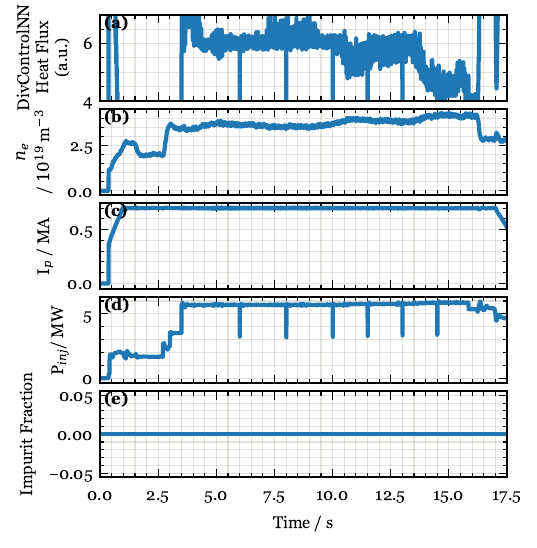}
 \caption{
KSTAR \#35854 DivControlNN quantities.
(a) Shows the calculated heat flux at outer divertor calculated by DivControlNN.
Rest of the panels are inputs to DivControlNN.
(b) Shows the line averaged electron density.
(c) Shows the plasma current.
(d) Shows the total injected power.
(e) Shows the impurity fraction estimate.
Note that this calculation malfunctioned and fed constant zero input to the model even though N$_2$ was puffed in this shot.
Apart from these inputs, diffusion scaling factor was set to a constant value of 1.0 in the model.
}
 \label{fig:SM_inputs_35854}
\end{figure}

\begin{figure}[!h]
 \centering
 \includegraphics[width=\linewidth]{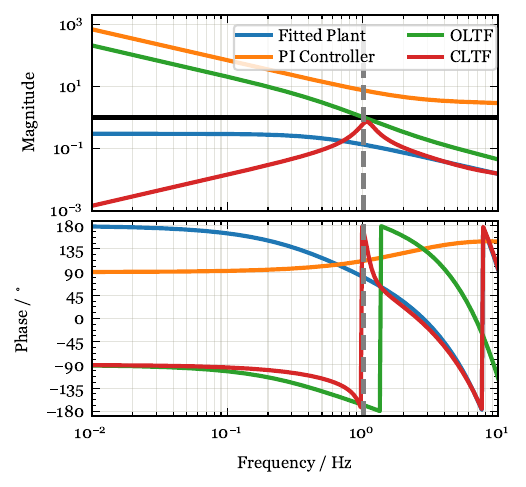}
 \caption{Closed loop transfer function analysis of the system using DivControlNN heat flux at outer divertor output with chosen PI controller with gains: \SMKp, \SMTi, and \SMstau.
The dashed grey vertical line shows the \ac{UGF} of the system.}
\label{fig:cltf_sm}
\end{figure}

Before we attempted detachment control experiments, we took two system identification shots.
The data from the first system identification shot \#35853 is shown in Fig.\ref{fig:sysid_afrac}.
In this shot, we puffed in N$_2$ gas in steps of 1.0 V, 2.5V, and 4.0 V with puff duration of 1.5s each.
A corresponding response was seen in \Afrac but with a delay.
We later confirmed from post-shot EFIT data that the strike point was indeed within the real-time Langmuir Probe array and thus our \Afrac calculation was valid.
We fitted the measured data with a simple first-order plant model (same as first order plus dead time\cite{Eldon_2022_PPCF}) of gain K, time constant $\tau$, and time delay $L$ given by (in Laplace domain):

\begin{equation}
 G(s) = \frac{K}{\tau s + 1}e^{-L s}
\label{eq:sysid}
\end{equation}

where $s$ is complex frequency variable in Laplace domain and $G(s)$ is the transfer function of the plant model.
The fit resulted in an identified model with \AfracK, \AfracTau, and \AfracL.
The fit is shown in Fig.\ref{fig:sysid_afrac}b.
Note that only the part of the time series data that was used in fit is shown for the fitted curve.
This fit was performed in the inter-shot interval during the experiment and has not been improved or modified after the experiment.

The controller gains were chosen by visualizing \ac{CLTF} of the system with chosen PI gains as shown in Fig.\ref{fig:cltf_afrac}.
Here, the frequency domain response of the plant model ($G(s)$) and PI controller ($T_{PI}(s)$) are plotted together.
When connected in series, this forms the \ac{OLTF} of the system ($O(s) = G(s) T_{PI}(s)$).
The frequency where \ac{OLTF} becomes 1.0 is called \ac{UGF}).
Phase margin is defined as the additional phase delay at \ac{UGF} that would make the system unstable by taking it to -180$^\circ$.
Additionally, we also define delay margin as the additional actuation delay that would make \ac{UGF} unstable.
The \ac{CLTF} is then calculated by solving the loop algebra in the Laplace domain:

\begin{equation}
 C(s) = \frac{G(s)}{1 + O(s)}
\label{eq:cltf}
\end{equation}

\begin{figure*}[!ht]
 \centering
 \includegraphics[width=\textwidth]{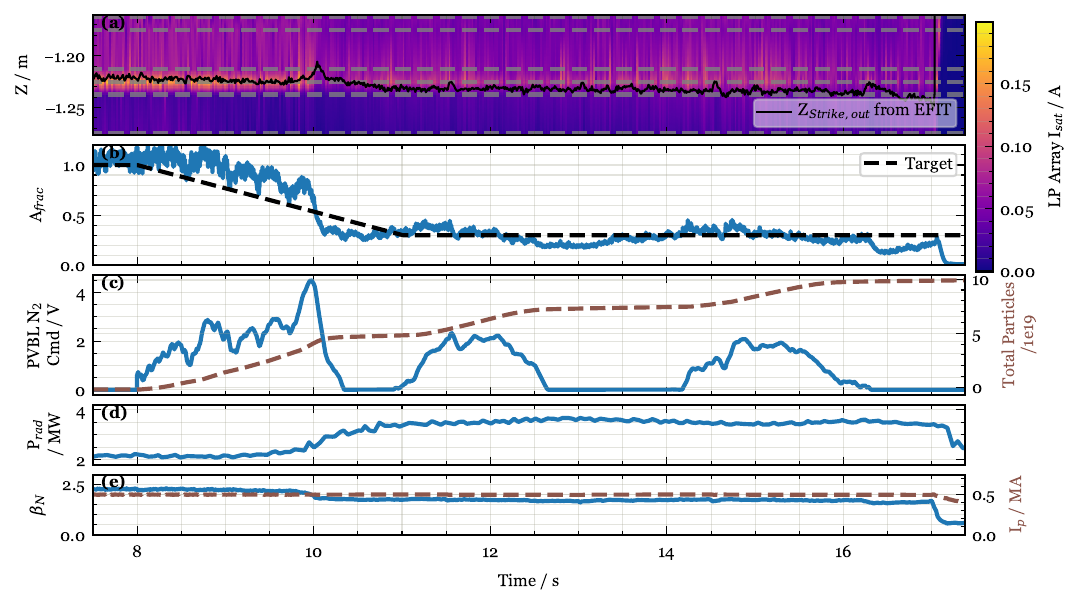}
 \caption{
Detachment control shot \#35857 using \Afrac controller.
(a) Shows the measured ion saturation current by realtime Langmuir Probe array at locations marked by grey dashed lines.
The data has been interpolated spatially using cubic spline interpolation.
The black curve shows the post-shot calculated strike point position on outer divertor using EFIT.
(b) Shows the \Afrac calculated from peak value among the Langmuir probe array.
The dashed black line shows the target provided to the controller to follow.
(c) Left axis: Shows the N$_2$ gas command steps sent for system identification.
Right axis: Shows the cummulative N$_2$ gas particles injected into the vessel.
(d) Total radiated power measured by \ac{IRVB}.
(e) Left axis: Shows $\beta_n$.
Right axis: Shows the plasma current (I$_p$).
}
 \label{fig:detctrl_afrac}
\end{figure*}

Because of the long delay, we chose to not use a derivative gain.
The goal of tuning was to push \ac{UGF} as high as possible (\AfracUGF) while keeping a reasonable phase margin (\AfracPhaseMargin) and margin for any additional actuation delay (\AfracDelayMargin).
This resulted in controller settings as: \AfracKp, \AfracTi, and \Afracstau, where K$_p$ is proportional gain, T$_i$ is integral time, and $\tau_s$ is pre-smoothing time constant.
This is still a very aggressive choice of controller, but given that the system identification fit gave an unexpectedly high value of response time $\tau=1$s probably due to too much noise during small step inputs, we decided to go ahead with this controller choice.
The resulting PI controller transfer function is given by:

\begin{equation}
 T_{PI}(s) = K_p \left( \frac{1}{T_i s} + 1\right) \frac{1}{1 + \tau_s s}
\label{eq:PI}
\end{equation}

Unfortunately, the surrogate model was not configured properly in this system identification shot due to technical errors, so we repeated a system identification but this time we decided to keep the nitrogen valve in the constant open position, to look for any deviation in the behavior.
The data from this second system identification shot is shown in Fig.\ref{fig:sysid_sm}.
Despite all the limitations of DivControlNN as listed earlier, we still saw a good correlation in the DivControlNN heat flux output at the outer divertor with the injected gas as seen in Fig.\ref{fig:sysid_sm}b.
This is validated by estimated \Afrac in Fig.\ref{fig:sysid_sm}d showing an increase in detachment level as the predicted output heat flux decreases.
The strike point was maintained within the realtime Langmuir probe array (Fig.\ref{fig:sysid_sm}a) validating the output of \Afrac.
Note that we have not yet calibrated the DivControlNN model with any experimental data, so we treated the output as arbitrary units and later attempted to control the detachment with estimated changes to this arbitrary output.

Fig.\ref{fig:SM_inputs_35854} shows the time-varying inputs to DivControlNN along with the heatflux output from the model.
Here, note that the impurity fraction input to the model malfunctioned, and a constant zero impurity fraction was fed to the model even though we puffed in N$_2$ in this system identification test. During the test part from 7.5~s onwards, we can see that most inputs to the model remained mostly constant, and only the line-averaged electron density showed considerable changes.
It can be seen thus that at this time, DivControlNN solely relied on changes in the electron density to estimate heat flux at the outer divertor.
The malfunctioning of the impurity fraction was detected in the post-processing of the data, and thus, during the experiment, we continued to try to use this model as it was.

We again fitted this system with a first-order system with a delay as described in Eq.\ref{eq:sysid}.
The fit resulted in an identified model with \SMK, \SMTau, and \SML.
The fit is shown in Fig.\ref{fig:sysid_sm}b.
Here as well, the fitting shown was performed during the experiment in the inter-shot interval and has not been modified or optimized later.
The time domain in which the fitting curve is shown is the data where the system was fitted.
Admittedly, this fit was not very good and we did not believe the large lag value to be accurate.
So for the purpose of tuning the controller, we arbitrarily set the system lag value to 100 ms.
The controller gains were chosen by visualizing \ac{CLTF} of the system with chosen PI gains as shown in Fig.\ref{fig:cltf_sm} and following the same procedure as we described for \Afrac controller tuning.
The resulting controller settings were: \SMKp, \SMTi, and \SMstau{} creating controller given by Eq.\ref{eq:PI}.
Here, we estimated to achieve a \ac{UGF} of \SMUGF, phase margin of \SMPhaseMargin, and delay margin of \SMDelayMargin.
This controller was also very aggressive, but we decided to go ahead with this controller choice given the limitations of the system identification fit and lack of time for further analysis in between the allotted run time of our experiment.
We understand that the system identification procedure is fitting a more complex and probably non-linear model with a simple linear model, and thus the above mentioned controller tuning technique is at best a good way to come up with an intial controller that can be adjusted better after the first closed loop test

\section{Results}
\label{sec:results}

\begin{figure*}[!ht]
 \centering
 \includegraphics[width=\textwidth]{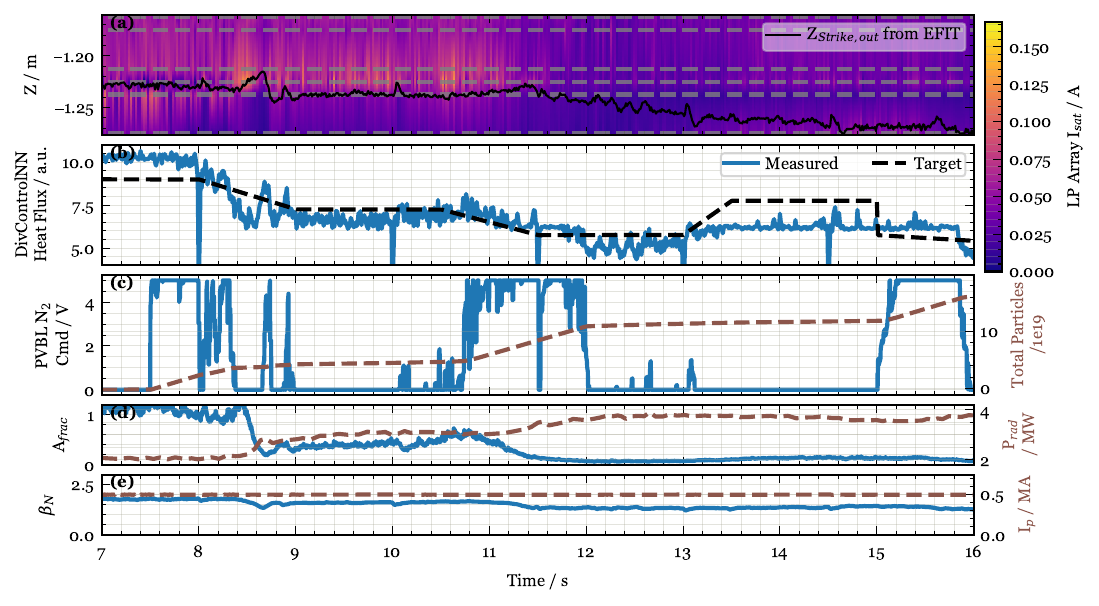}
 \caption{
Detachment control shot \#36161 using DivControlNN heat flux at outer divertor.
(a) Shows the measured ion saturation current by realtime Langmuir Probe array at locations marked by grey dashed lines.
The data has been interpolated spatially using cubic spline interpolation.
The black curve shows the post-shot calculated strike point position on outer divertor using EFIT.
(b) Shows the heat flux at outer divertor calculated by DivControlNN.
The dashed black line shows the target provided to the controller to follow.
(c) Left axis: Shows the N$_2$ gas command steps sent for system identification.
Right axis: Shows the cummulative N$_2$ gas particles injected into the vessel.
(d) Left axis: Shows the \Afrac calculated from peak value among the Langmuir probe array.
Right axis: Total radiated power measured by \ac{IRVB}.
(e) Left axis: Shows $\beta_n$.
Right axis: Shows the plasma current (I$_p$).
}
 \label{fig:detctrl_sm}
\end{figure*}

\begin{figure}[!h]
 \centering
 \includegraphics[width=\linewidth]{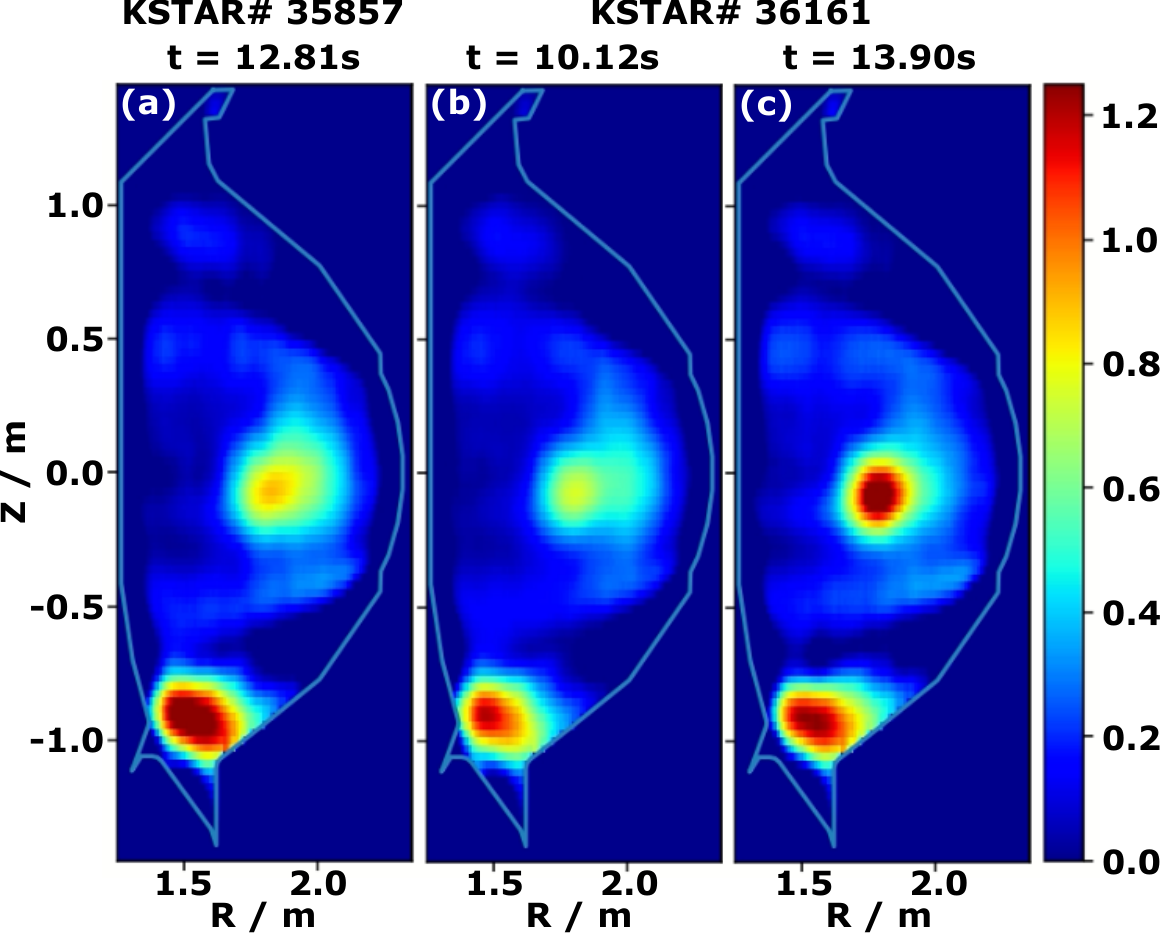}
 \caption{
KSTAR \ac{IRVB} measured radiated power density (a.u.) inverted into 2D cross-section.
(a) KSTAR \#35857 at 12.81s at the peak of total radiated power.
(b) KSTAR \#36161 at 10.12s before  the second impulse of gas between 10.8s to 12s.
(c) KSTAR \#36161 at 13.90 after the last gas impulse.
}
\label{fig:prad_2d}
\end{figure}

Utilizing the controllers tuned in Sec.\ref{sec:sysid}, we attempted detachment control experiments.
First, we used \Afrac controller in KSTAR \#35857 with results shown in Fig.\ref{fig:detctrl_afrac}.
As can be seen in Fig.\ref{fig:detctrl_afrac}a, the strike point remained within 12.5~mm of one of the probes of the realtime Langmuir probe array, thus the obtained \Afrac signal shown in Fig.\ref{fig:detctrl_afrac}b has error $\lesssim$13\%.
Here, we can see that the controller was successful in closely following the target provided to it completing the pre-programmed shot length to the end.
It is also evident that the aggressive control strategy was good and did not result in any long sustained oscillations while providing a quick response to the changing target value.
From 8s to 10s, it can be seen that the injected N$_2$ was just enough to ramp down the measured \Afrac with the same slope.
The accumulated offset from the target eventually caused the integral term to send a brief impulse of nitrogen near 9.8~s and then the controller further converged with the target value.
For the rest of the shot, small nitrogen puffs were required to correct the drifting \Afrac and keep it on the target.
The total radiated power from the plasma as measured by KSTAR \ac{IRVB} remained below 3.5 MW ($\approx$65\% of injected power) and as can be seen in Fig.\ref{fig:prad_2d}a (snapshot taken at around the time of maximum radiation), the majority of the radiation was coming from the divertor region and the core region was not losing power through radiation.
This further validated, that \Afrac controller strategy first devised in Ref.\cite{Eldon_2022_PPCF} is a viable option for detachment control even with Tungsten divertor in KSTAR.

Since \Afrac controller has been demonstrated in the past as well, we decided to utilize the remaining allotted runtime on KSTAR to test the DivControlNN prototype-based controller.
Fig.\ref{fig:detctrl_sm} shows the results from shot \#36161 where we deployed this controller.
An immediate issue was seen with DivControlNN output that the initial heat flux calculation had a different starting value than what we saw in reference shots and system identification shot \#35854.
Because of this, when the controller turned on at 7.5s, the large error resulted in the railing of gas command output which caused too much N$_2$ injected into the system.
While this quickly brought down the measured signal, it also resulted in an overshoot.
In the next ramp-down of the target from 10.5s to 11.5s, more impurity was injected as we tuned an aggressive controller.
It can be seen from \Afrac in Fig.\ref{fig:detctrl_sm}d that the system reached deep detachment by this point and the ion saturation current measurements (Fig.\ref{fig:detctrl_sm}a) became unreliable beyond 12~s.
In post-analysis of IRVB inverted data as shown in Fig.\ref{fig:prad_2d}, we can see that the core started radiating a lot of power after the last railed impulse of gas input between 10.8~s to 12~s.
This would have reduced $P_{SOL}$ (power entering SOL from core) and started to starve the divertor of power, driving or contributing to the failure to follow the rising heat flux target shown in Fig.\ref{fig:detctrl_sm}b between 13 and 15~s.
There is no surprise with the fact that setting target values for an uncalibrated output is not always deterministic and would cause issues as we faced.

Post-shot data analysis discovered further issues in our operation of the DivControlNN model.
The impurity fraction calculation as mentioned in Sec.\ref{sec:control_variables} malfunctioned and sent a constant zero input to the model (see Fig.\ref{fig:SM_inputs_35854}e).
Thus the model was unable to respond directly to large amounts of impurity that were injected into the system and was only relying on data from line-integrated core electron density, input power, and plasma current as realtime inputs.
Despite these limitations, this preliminary test sheds light on the potential of using such a surrogate model-based controller for detachment control in future reactors.

\section{Discussion}
\label{sec:discussion}

In this paper, we described re-using \Afrac as a reliable control variable for detachment control provided that realtime ion saturation current measurements are available from the Langmuir probes and the strike point is controlled well enough that such an array can be used to estimate peak $I_{sat}$.
It can be seen from panels (a) and (c) in Fig.\ref{fig:sysid_afrac}, Fig.\ref{fig:sysid_sm}, and Fig.\ref{fig:detctrl_sm} that when the total injected impurity amount crosses a rough threshold of about $1\times10^{20}$ particles, the plasma boundary shape is deformed such that strike point on outer divertor starts drifting downwards (and the inner strike point moves upwards) even though the X-point is held in place by the magnetic shape control system.
Thus, \Afrac controller is best suited with an outer strike point control system commissioned on the device, as was demonstrated in the previous carbon divertor \cite{Eldon_2022_PPCF}.
Unfortunately, direct strike point position control (rather than X-point position control), was not yet commissioned for the new divertor at the time of these tests.
It is also important to keep note of the position of strike point and the width of the ion saturation current profile on the divertor.
In our case, we estimated that the ion saturation current profile width is about 16~mm, just enough to ensure that at least one of the Langmuir probes is always inside the wetted area from the \ac{SOL} plasma.
Even then, it can be seen that at around 12.4s in Fig.\ref{fig:sysid_afrac} and at around 10s in Fig.\ref{fig:detctrl_afrac} that as the strike point moves from OD9 (Z=-1.225m) to OD11(Z=-1.2125m), the corresponding \Afrac value shows a sharp decline and then recovery, probably due to the peak passing through the middle of the two probes.
This effect is small enough that our existing controller was able to fix it, but it shows a potential source of error in system identification and might also cause loss of control if the sudden transition can excite an unstable oscillation of the \ac{UGF}.
For future applications of this controller, we are working on including realtime spatial analysis of the $J_{sat}$ profile, potentially informed with profile shapes from high-fidelity simulations from SOLPS-ITER or UEDGE.

In our experimental session, since strike point control had not been commissioned, we attempted real-time empirical profile analysis with strike point sweeps.
But we found that the actuation strength and response time of the poloidal field coils at KSTAR do not allow for large enough and fast enough strike point sweeps.

It should be noted that in the application of \Afrac controller method on KSTAR, tuning the overall factor to \Afrac so that it reports 1.0 when fully attached was trickier than the case for full carbon environment KSTAR\cite{Eldon_2022_PPCF}.
We noticed offsets in the outputs of Langmuir probes which changed from shot to shot, and thus ensuring the correct normalizing factor for \Afrac became harder.
This was the reason why we had to change the factor for \Afrac after shot \#35853 as also mentioned in the caption of Fig.\ref{fig:sysid_afrac}.
After this experience, we have now added an online offset estimator and subtraction for all probe signals, which measures the offset before the plasma breakdown and ensures that the zero offset is correct on the probes.
This issue is likely due to electrical connectivity problems with the probes which also showed other issues during the campaign, but still, this experience should be noted for future reproduction and improvements.

Another point of uncertainty in \Afrac model could be the magnetic connection length between the upstream (outboard midplane) and divertor in the 2PM\cite{Leonard_2018_PPCF}. This length is kept fixed in the model and while we did not change the plasma boundary shape much from the previous test\cite{Eldon_2022_PPCF} in carbon divertor KSTAR, it still makes it a potential source of error in wider usage in future. The realtime equilibirum calculations during shot provided by RTEFIT is being upgraded to also output this magnetic connection length so that in future the model gets more accurate and time varying information about this important parameter.

Although Langmuir probes might not be able to survive future burning plasma experiments, they are still a valuable tool for studying detachment control experiments for ease of installation and operation in experimental devices.
Even in burning plasma devices, sacrificial Langmuir probes can be used in initial device commissioning and preparation of base scenarios at low power.
Knowing what a stable detached scenario should be like could significantly decrease the time required to commission controllers based on other control variables, and give a baseline level of detachment control performance to compare them to.
Thus, Langmuir probe based control might provide a foothold in future device commissioning, as it has been shown to be useful on many devices \cite{Eldon_2021_NME, Guillemaut_2017_PPCF, Yuan_2020_FED, Khodunov_2021_PPCF}.
The good results from \Afrac controller as seen in Fig.\ref{fig:detctrl_afrac} could also motivate further research in similar biased electrode measurement methods of SOL plasma such as biased divertor plates \cite{Toi_2023_NF, Cui_2024_NF} which behave like larger area Langmuir probes and can withstand harsher conditions in comparison to small tip area probes.

We also demonstrated using a machine-learning-based surrogate model, DivControlNN, which infers from a large database of 2D UEDGE simulations for estimating hard to infer quantities in the plasma, such as heat flux on the divertor, for controlling detachment level with realtime feedback.
As of the writing of this manuscript, this detachment control method is the first of its kind ever implemented and will act as a stepping stone for future deployments.
This is an important step in the direction of achieving detachment control in future burning plasma reactors which would have very limited means of measuring the detachment level due to space constraints and harsh environment.

We have identified critical weak points in the prototype of DivControlNN and the control infrastructure required to utilize this model, and we are working on improving these aspects for future tests.
One likely mistake we made during the experiments was treating the long dead time reported by DivControlNN heatflux output in response to gas puff (Fig.\ref{fig:sysid_sm}) as an overstimate.
Since DivControlNN had a constant zero impuirty fraction concentration, it was solely relying on line integrated core electron density information for responding to changes.
While ion saturation currents provide local divertor information fast, the core electron density response to gas puff would have additional transport timescales and thus DivControlNN output might truly have a larger daed time.
This could have resulted in the high controller gain that saturated the gas response in the test (Fig.\ref{fig:detctrl_sm}).

We are in the process of creating a new 2D UEDGE database of KSTAR with a tungsten divertor and considering multiple charged states of additional impurities such as nitrogen, neon, and argon.
New models would be trained on the expanded database and acquired experimental data from this campaign, with the input of injected gas flow instead of impurity fraction to simplify the use case of these models.
We would also work with the KSTAR team to improve PCS communication infrastructure so that accurate realtime values of injected power are available to our models.

The initial success of the neural network surrogate model in detachment control motivates and corroborates similar studies, simulations, and training of other models for providing fast estimates of plasma parameters, for quick decision-making in the control room during experiments, as well as, for potential use in other control systems where important plasma properties are often not accessible directly.
A neural network based control system approach has already been demonstrated in magnetic shape control\cite{Degrave_2022_Nature}.
For \ac{SOL} plasma predictions, machine learning surrogate models were first pioneered using 1D UEDGE simulations\cite{Zhu_2022_JPP}, serving as the proof-of-principle study that paved the way for DivControlNN presented here, which is based on 2D UEDGE simulations.
More recently, model based on Hermes-3\cite{Dudson_2024_CPC} simulations of MAST-U\cite{Holt_2024_NF} and neural partial differential equation solver for TCV\cite{Poels_2023_NF} have been reported and are under further development.

Another major focus of future experiments would be to use noble gases in detachment control.
N$_2$, while being excellent at cooling the SOL plasma in conventional tokamaks, would not be a good impurity to seed in tritium fueled plasma due to the formation of tritiated ammonia\cite{Pitts_2019_NME}.
Such tritiated ammonia would require additonal tritium reclaiming processes which would reduce the duty cycle of reactors, as well as, pose additional risks in handling a radioactive gas.
We are in the process of testing Ne and Ar as alternate cooling gases.
In the KSTAR scenarios investigated so far, the effect of Ne on detachment has been hard to observe as small gas puffs do not actuate enough on the \ac{SOL} plasma but if the gas puffing is increased, we suddenly observe disruption due to too much cooling inside the separatrix.

\section*{CRediT authorship contribution statement}

\textbf{A.~Gupta} investigation, methodology, software, formal analysis, visualization, writing - original draft.
\textbf{D.~Eldon} conceptualization, software, investigation, writing - review \& editing, supervision, funding acquisition.
\textbf{E.~Bang} resources, data curation
\textbf{K.~Kwon} resources, investigation
\textbf{H.~Lee} resources, supervision
\textbf{A.~Leonard} writing - review \& editing
\textbf{J.~Hwang} resources, data curation
\textbf{X.~Xu} conceptualization, supervision, funding acquisition
\textbf{M.~Zhao} methodology, software
\textbf{B.~Zhu} methodology, software

\section*{Declaration of competing interest}

The authors declare that they have no known competing financial interests or personal relationships that could have appeared to influence the work reported in this paper.

\section*{Acknowledgements}

This material is based upon work supported by the U.S. Department of Energy, Office of Science, Office of Fusion Energy Sciences,
under Awards 
DE-SC0023400, 
and
DE-AC52-07NA27344.  

This research was supported by R\&D Program of ``Korea-US Collaboration Research for High Performance Plasma on Tungsten Divertor(EN2503)" through the Korea Institute of Fusion Energy(KFE) funded by the Government funds, Republic of Korea.

\section*{Disclaimer}

This report was prepared as an account of work sponsored by an agency of the United States Government.
Neither the United States Government nor any agency thereof, nor any of their employees, makes any warranty, express or implied, or assumes any legal liability or responsibility for the accuracy, completeness, or usefulness of any information, apparatus, product, or process disclosed, or represents that its use would not infringe privately owned rights. 
Reference herein to any specific commercial product, process, or service by trade name, trademark, manufacturer, or otherwise, does not necessarily constitute or imply its endorsement, recommendation, or favoring by the United States Government or any agency thereof.
The views and opinions of authors expressed herein do not necessarily state or reflect those of the United States Government or any agency thereof.

\bibliography{refs}

\end{document}